\documentclass{ws-p9-75x6-50}

\begin{document}

\title{Quantum phases in two-dimensional
 frustrated spin-1/2 antiferromagnets}

\author{P. Sindzingre}

\address{Laboratoire de Physique Th\'eorique des Liquides-UMR 7600 of CNRS,
Universit\'e  Pierre  et  Marie  Curie,  case 121,
4 place Jussieu, 75252 Paris Cedex, France 
\\E-mail: phsi@lptl.jussieu.fr}

\author{C. Lhuillier, J.-B. Fouet}

% \address{Another address}

%%%%%%%%%%%%%%%%%%%%%%%%%%%%%%%%%%%%%%%%%%%%%%%%%%%%%%%%%%%%%%
% You may repeat \author \address as often as necessary      %
%%%%%%%%%%%%%%%%%%%%%%%%%%%%%%%%%%%%%%%%%%%%%%%%%%%%%%%%%%%%%%

\maketitle

\abstracts{
We describe four phases found in two-dimensional quantum antiferromagnets.
Two of them display long range order at $T=0$: the N\'eel state and the
Valence Bond Crystal. The last two are Spin-Liquids. Properties of these
different states are shortly described and likely conditions of their 
occurence outlined.
}

\section{INTRODUCTION}
Investigating various spin-1/2 systems in two dimensions (2d),
by means of exact diagonalization (ED) on small samples of $N$  spins,
we have found, up to now, four kinds of antiferromagnetic
phases at T=0. Two of them display long range order (LRO):
N\'eel or Valence Bond Crystal (VBC) LRO. The others are two different
Spin-Liquids. The N\'eel case may be qualified as semi-classical:
it is a ground-state of the (spin-$\infty$) classical model
with an order parameter reduced by quantum fluctuations.
The others are purely quantum phases.

The hamiltonians considered are $SU(2)$ invariant,
like the Heisenberg hamiltonian and
its generalizations as the $J_1-J_2$ model or the $J_1-J_2-J_3$ model
and the multiple-spin exchange (MSE) model.
All these models of magnetism picture exchange (or super-exchange)
between fermions.
The permutation operator of two spins reads:
\begin{equation}
P_{ij}= 2  {\bf S}_i \cdot {\bf S}_j + {1\over 2}
\label{eq:exchange}
\end{equation}   
This yields  the Heisenberg hamiltonian:
\begin{equation}
H=   \sum_{<ij>} {\bf S}_i \cdot {\bf S}_j
\label{eq:heisen}
\end{equation}
where the sum runs over nearest neighbor pairs of spins on sites $i$ and $j$,
the $J_1-J_2$ model:   
\begin{equation}
H=   J_1 \sum_{<ij>} {\bf S}_i \cdot {\bf S}_j
 +   J_2 \sum_{<<ij>>} {\bf S}_i \cdot {\bf S}_j
\label{eq:j1j2}
\end{equation}
where the second sum runs over next-nearest neighbor pairs
and the $J_1-J_2-J_3$ model with additional 3th nearest neighbor
interrations.
We were also motivated to study MSE models involving exchange of more 
than two spins. Among them we shall focus on the simplest
which brings new physics which involves up to 4-spin exchange.
Its hamiltonian reads:
\begin{equation}
H=  J_2  \sum_{<ij>}  {P}_{ij} 
   +  J_4 \sum_{<ijkl>} ( {P}_{ijkl}  + {P}_{ijkl}^{-1} )
\label{eq:mse}
\end{equation}
where ${P}_{ijkl}$ stands for the permutation of 4 spins on sites $i,j,k,l$,
${P}_{ijkl}^{-1}$ is the inverse permutation
(note that ${P}_{ij}={P}_{ij}^{-1}$).
The 2d lattices are the square, honeycomb, triangular and kagom\'e lattices.

\begin{figure}[t]
% \figurebox{22pc}{15pc}{} % to have a box alone
 \vspace{-8pc} 
 \hspace{-3pc}
 \epsfxsize=20pc
 \epsfbox{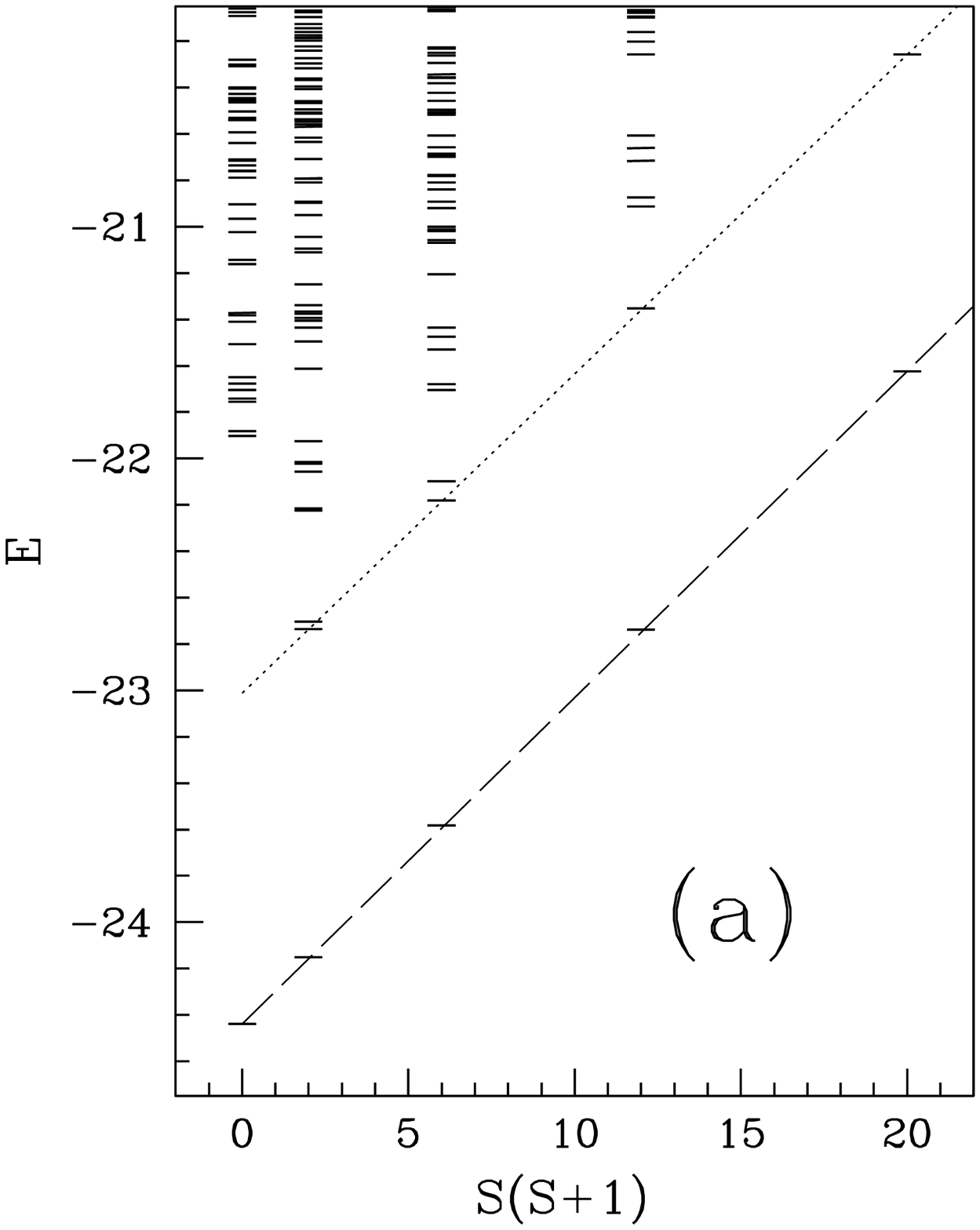}
 \vspace{-20pc} 

 \hspace{13.5pc}
 \epsfxsize=20pc
 \epsfbox{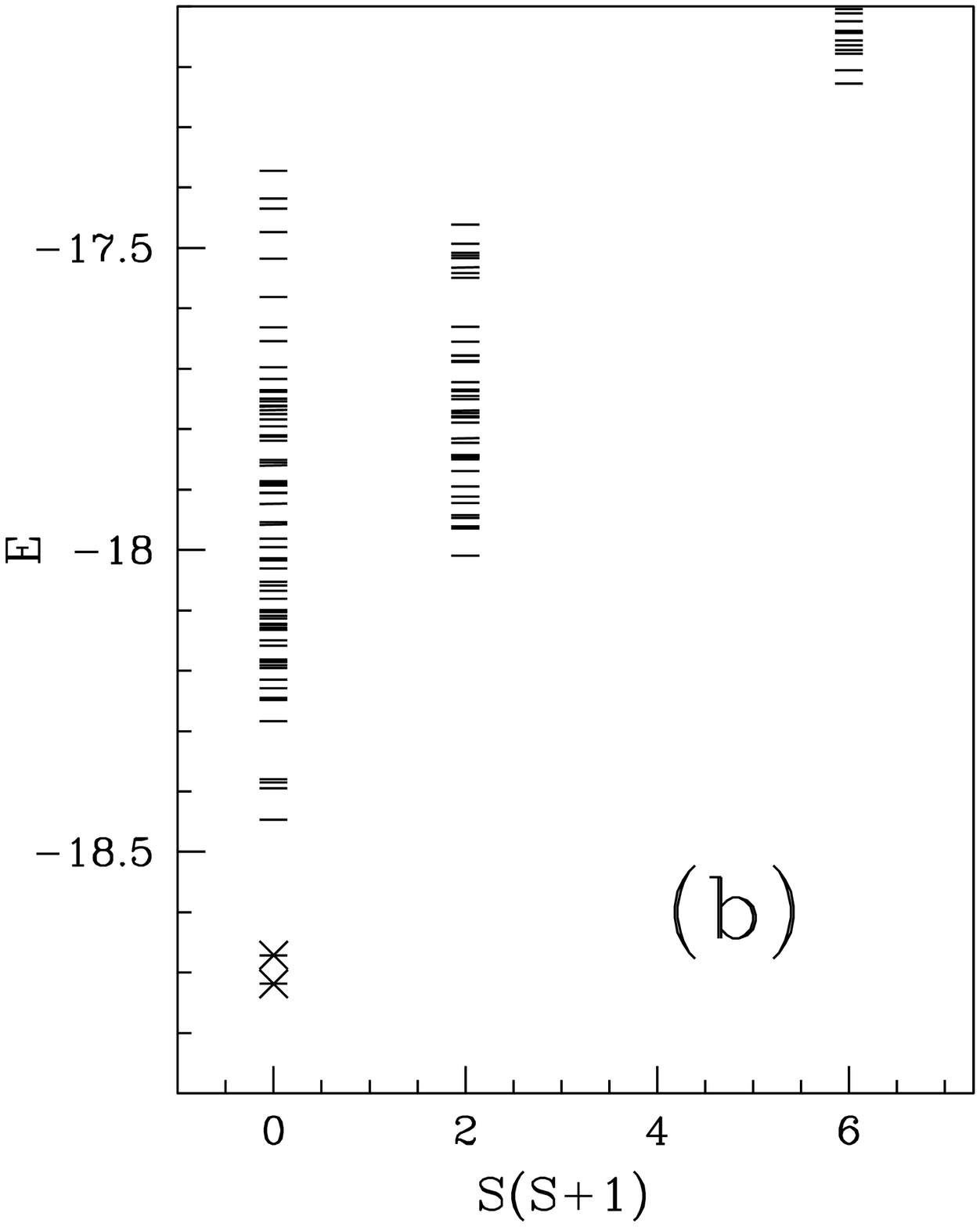}
 \vspace{-1pc} 

 \epsfxsize=20pc
 \hspace{-3pc}
 \epsfbox{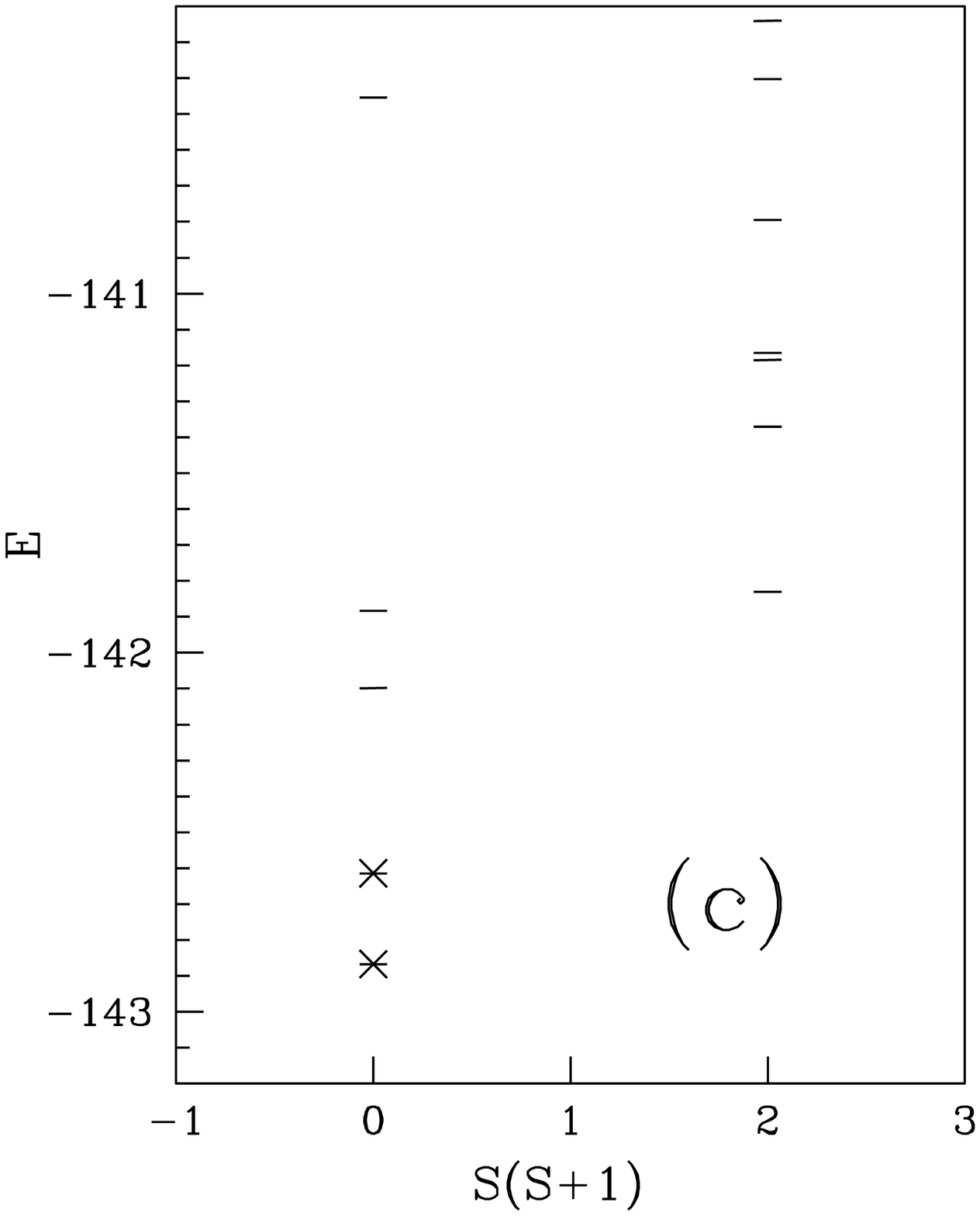} 
 \vspace{-20pc} 

\hspace{13.5pc}
 \epsfxsize=20pc
\epsfbox{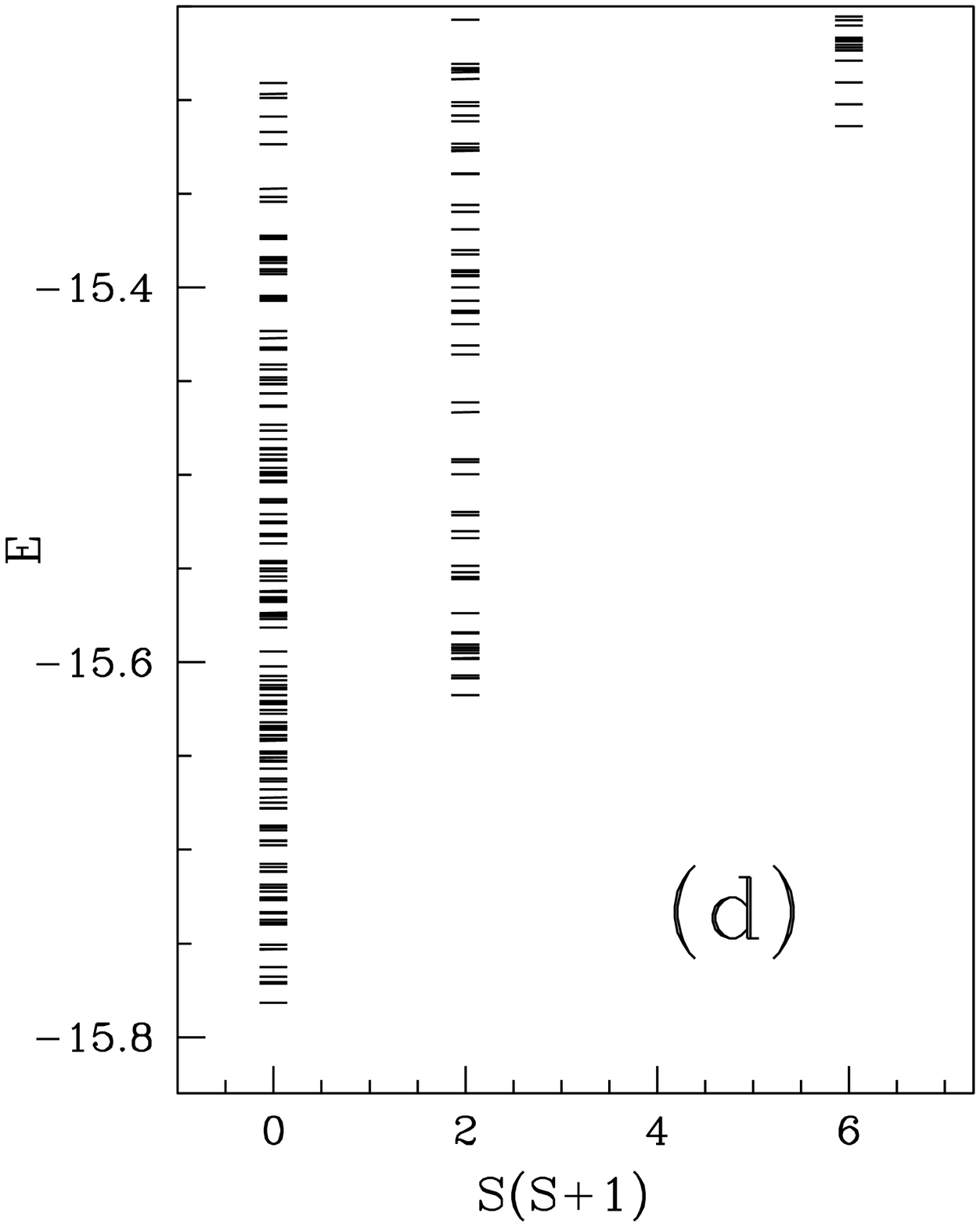} 
\caption{Eigen-energies vs $S(S+1)$ of $N=36$ samples.
(a) Spectrum of the Heisenberg model on the square lattice,
characteristic of collinear N\'eel LRO:
the lowest energies in each spin sector $S$ increase as $S(S+1)$
with $S$ (states on the dashed line) and are well separated from
the softest magnons (states on the dotted line).
(b) Spectrum of the Heisenberg model on the checkerboard lattice,
which displays VBC LRO:
for $N\rightarrow\infty$, the two lowest $S=0$ states (crosses)
become degenerate
and the gap to the others states remains finite,
the finite spin-gap implies that $SU(2)$ remains unbroken.
(c) Spectrum of the MSE model (eq.~\ref{eq:mse} with $J_2=-2$, $J_4=1$)
on the triangular lattice, a Type I Spin Liquid:
four $S=0$ states (crosses) become degenerate as $N\rightarrow\infty$,
the excitations above the degenerate ground-state remain gapped.
(d) Spectrum of the Heisenberg model on the  kagom\'e lattice,
a Type II Spin Liquid: there is
a spin-gap and a gapless continuum of singlet states.
}
\label{fig:1}
\end{figure}

In order to distinguish between the different phases of such models,
from ED calculations  on small samples
(of up to $N=36$ spins with present day computers),
the analysis of the symmetries and scaling properties of the low-lying
eigen-levels of the spectra is a very efficient approach.
Typical spectra of systems in the four phases identified are displayed
in Fig~\ref{fig:1}
as a function of the eigenvalues $S(S+1)$ of the square ${\bf S}^2$ of the
total spin (a good quantum number as the hamiltonians are $SU(2)$ invariant).
A superficial analysis allows an immediate detection of obvious distinctive
features:
\begin{itemize}
\item  Fig~\ref{fig:1}a: N\'eel spectrum, the lowest eigen-levels in
each $S$ sector form a separate set with an energy increasing as $S(S+1)$
(the quantum origin of the Landau free energy per spin $f$ increasing
with the magnetization $m=S/(N/2)$ as
$f(m)=f(0)+m^{2}/2\chi+\cdots$ for small $m$,
where $\chi$ is the spin susceptibility).
\item Fig~\ref{fig:1}b and ~\ref{fig:1}c: VBC or type I Spin-Liquid spectra
with a gap $\Delta$ between different sector of spins
($f(m)= f(0)+ \Delta m/2 +\cdots$)
\item Fig~\ref{fig:1}d: type II Spin-Liquid spectrum with a spin-gap
and a gapless continuum of singlets excitations
\end{itemize}
We now briefly sketch the spectral properties of the four different phases
and outline the groups of systems where they have been found.

\section{Semi-classical N\'eel phase}
N\'eel LRO breaks $SU(2)$ 
and various space symmetries at the thermodynamic limit.  
This is revealed in the spectra by a basic set of low-lying eigen-levels 
(states on the dashed line in Fig~\ref{fig:1}a),
called QDJS for Quasi Degenerate Joint States in Refs.
\cite{blp92,bllp94}, with energies $\sim S(S+1)/2N\chi$ 
at least for $S$ values up to $\sqrt{N}$.
They collapse together to the absolute ground-state as $1/N$
and separate from the softest magnons (dotted line in Fig~\ref{fig:1}a)
which decrease as $1/\sqrt{N}$.  The QDJS have different $S$ values
which enable breaking of $SU(2)$ symmetry. The number of QDJS per $S$ value
is caracteristic: it is one for collinear LRO and equal to the number of way
of adding $p$ spins of size $N/2p$ for noncollinear LRO with $p$ sublattices.
The QDJS belong to different irreducible representations of the
space group of the lattice which allow and reveal the space symmetry breaking
occuring. For instance, in case of the Heisenberg model on the 
square lattice of Fig~\ref{fig:1}a, where translation symmetry is broken,
the QDJS have wave vectors ${\bf k}= (0,0)$ for even $S$
and  ${\bf k}= (\pi,\pi)$ for odd $S$ values and are invariant under
all operations of the point group (taking the origine on a site),
as expected for this collinear N\'eel LRO.
ED and/or Monte-Carlo calculations have shown that
the Heisenberg model on the 2d bipartite lattices 
investigated up to now:
square (see Refs.\cite{sz92,sandvik-97,beard-wiese-96} and 
references therein),
honeycomb\cite{rry89,fsl00},
1/5 depleted square\cite{tku-96}$\cdots$, 
displays collinear N\'eel LRO. 
In these unfrustrated systems, Monte-Carlo methods can deal now 
with very large systems which enable 
the investigation of properties out of range of 
the ED approach\cite{sandvik-97,beard-wiese-96}. 
Monte-Carlo approaches, however, is limited by the well-known
sign problem\cite{henelius-sandvik-01,cbps01}
in frustrated systems like those considered below.

A similar analysis  of the ED spectra
has shown that the Heisenberg model on the triangular lattice
displays noncollinear N\'eel LRO\cite{blp92,bllp94}.
It also revealed occurences of "order by disorder"
in the $J_1-J_2$ model on the triangular lattice\cite{lblp95}
and the $J_1-J_2-J_3$ model on the honeycomb lattice\cite{fsl00}.
In these situations the classical ground-state is a degenerate 
manifold of different kinds of 4-sublattice N\'eel order.
Evidence of the selection of a collinear LRO by quantum fluctuations
appears when increasing $N$, with the separation of the QDJS
specific of collinear LRO from the states associated with 4-sublattice
order.

Calculations of spin-spin correlations, order parameters,
spin susceptibilities \cite{bllp94}, stiffnesses\cite{lblp95a}
yield results consistent with the spectrum analysis but
their finite scaling behavior is less favourable,
(i.e. $\sim 1/\sqrt{N}$ in N\'eel systems) 
and extrapolation to $N\rightarrow\infty$  could be rather inaccurate.
Agreement between the scaling behaviors of ED and spin wave results 
can nevertheless help to circomvent this limitation.

\section{VBC LRO}
This class includes systems with LRO of singlet objects such as
singlet bonds (dimers) or more extented ones (4-site plaquettes...).
This kind of LRO  usually breaks the space symmetry, so the 
spectrum must display a corresponding degeneracy in the thermodynamic limit,
but it does not break $SU(2)$ which manifest by a finite spin-gap.
Such features are visible in Fig~\ref{fig:1}b
which displays the spectrum of the Heisenberg model
on the checkerboard lattice\cite{fmsl01}: 
two $S=0$ states appear in the bottom of the spectrum well separated
from all others.
A finite size analysis confirm the degeneracy of the two states and 
that the singlet and the spin gaps remain finite for $N\rightarrow\infty$.
In the present case,
symmetry properties of the two states indicate unambigously LRO of 
plaquettes on the void squares of the checkerboard lattice,
which is confirmed by calculations of correlation functions.
Since the excitations are separated from the degenerate ground-state
by a gap $\Delta$, the thermal  behavior of the
spin susceptibility $\chi$ and heat capacity $C_v$ is 
exponentially activated $\sim {\rm exp}(-\Delta/kT)$ at low $T$.
Due to the gap $\Delta$, the ground-state energy in the different 
spin sector $S$ evolves as $E(S)\approx E_0+\Delta S+\cdots$ at low $S$.
This leads, in a continous Landau description
 to $f(m)\approx f(0)+\Delta m/2 -mh/2 +\cdots$
in the presence of a magnetic field $h$:
a finite field $h_c=\Delta$ is needed to induce a transition to states with
a non zero magnetization,
the magnetization curve $m(h)$ starts with a plateau at zero $m$.

Much attention has been given to the $J_1-J_2$ model on the square lattice
which for $J_2/J_1\sim0.5$ has been suspected to display 
dimer or plaquette LRO, but the issue is still debated
(see\cite{cls00,cbps01} and references therein).
Besides the checkerboard lattice\cite{fmsl01},
we have found a VBC phase in the $J_1-J_2$ model on the honeycomb 
lattice\cite{fsl00}.
In these three systems the VBC phase is located in the phase diagram near
a collinear N\'eel phase.
VBC LRO appears as a generic situation to be 
expected after destabilization by quantum fluctuations 
of a collinear N\'eel phase with increased frustration.
This is in agreement with various 
analytical calculations, based on
large-$N$ $SU(N)$ or $Sp(N)$ expansions
or non linear sigma-model theory
taking into account the Berry phases 
effects\cite{rs90,read-sachdev91,sachdev-park01}.
But ED calculations, limited to small systems, does not allow to investigate
the phase transition between N\'eel and VBC LRO
(see Refs\cite{sachdev-park01,sow01} for recently proposed scenarios).
Experimental evidence of VBC LRO has been found in 
the 2d insulating compounds:
$SrCu_2(BO_3)_2$\cite{kageyama99}
and $CaV_4O_9$\cite{taniguchi-95}.
These systems are however somewhat special.
The 2d lattices of magnetic ions in these compounds  have lower
symmetries than the above-mentioned lattices.
The hamiltonian already favors the formation of singlet objects.
There is no further symmetry breaking.

\section{Type I Spin Liquid}
The spectrum of type I Spin Liquid (see Fig~\ref{fig:1}c)
seems at first sight very similar to the spectrum of a VBC.
It has a a finite gap above a degenerate ground-state.
So $\chi$ and $C_v$ are thermally activated
and $SU(2)$ is not broken.
But we are lead to exclude VBC LRO because
the few $S=0$ levels collapsing together to the ground-state do not have 
symmetries that could be explained  in the framework a VBC picture
and all correlation functions are found short-ranged.
Indeed, in the two systems where we think that 
type I Spin Liquid behavior occurs: for the MSE model
on the triangular lattice\cite{mblw98,mlbw99}
and in the $J_1-J_2-J_3$ model on the honeycomb lattice\cite{fsl00},
VBC LRO would require a larger
degeneracy  of the ground-state than the one observed. 
For instance, the simplest VBC LRO on the triangular lattice is
12-fold degenerate whereas the observed degeneracy is 4-fold.
This 4-fold degeneracy is  not the signature of LRO
in a local order parameter but has a
topological origin\cite{ml00-lsmtheorem,sachdev-park01}.

Several aspects of the physics of type I Spin Liquids,
still deserve further investigations.
This fully gapped phase without VBC LRO is likely to be described
with short ranged resonating valence bonds (RVB).
A possible connection with the RVB phase of quantum dimer model,
such as the one found on the triangular lattice\cite{moessner-sondhi-01}
is worth to be clarified.
In type I Spin Liquids, the excitations are expected to consist of
deconfined spinons\cite{sachdev-park01,sachdev92}.
We are presently searching for numerical evidence of deconfined spinons.
It also remains to firmly outline
the conditions of appearance of type I Spin Liquids.
Fully gapped Spin Liquids are predicted to occur after
destabilization of non-collinear LRO\cite{sachdev-park01,sachdev92}.
The two examples of type I Spin Liquid, found up to now,
do not disagree with this. But, as discussed below,
type II Spin Liquid behavior with gapless excitations 
(not described in the 
approach of Refs\cite{sachdev-park01,sachdev92}),
may appear after destabilization of non-collinear LRO.
A further clue might be of importance:
in the two type I Spin Liquids short range
correlations are slightly ferromagnetic. In fact, they
appears in the vicinity of a ferromagnetic phase.

Experimentally, this phase might be observed in the low density
2d layers of solid $^3He$ where there is
evidence that the MSE model on the triangular lattice
with coupling constants that might be in the good range,
is realized
(see \cite{mlbw99,rbbcg97,bbccgrs00,imyf97} and references therein)
and possibly in the Wigner crystal\cite{bc00}.
It might also occur in the
${BaCo_2(AsO_4)_2}$ compound\cite{regnault-1},
likely to be described by the $J_1-J_2-J_3$ model on the honeycomb lattice
with parameters close to those for which we found indications of this
behavior. 

\section{Type II Spin Liquid }
\begin{figure}[t]
 \vspace{-2pc} 
 \hspace{2pc} 
 \epsfxsize=22pc 
\epsfbox{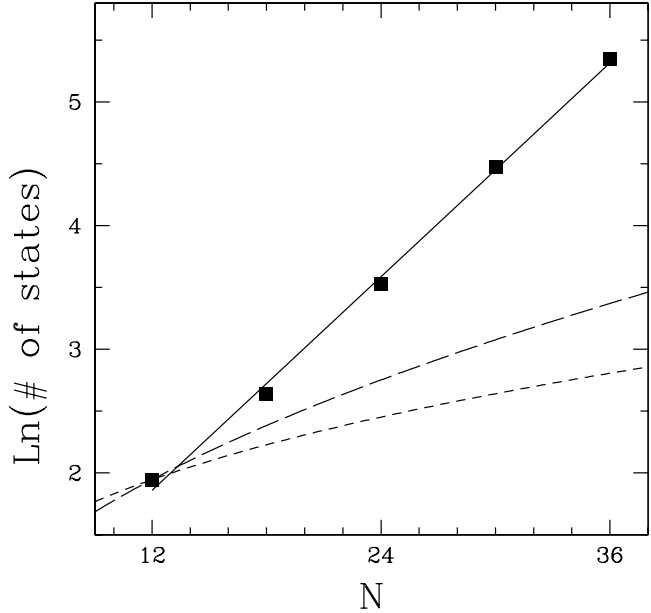} % postscript image file name
\caption{Heisenberg model on the kagom\'e lattice. Logarithm of the
number of states in the spin-gap (squares fitted by the full line)
vs size $N$ of the samples.
The observed behavior differs from the scaling law 
${\rm exp}(bN^{n/(n+2)})$ deduced from a single mode description of the
continuum with a dispersion law
$\epsilon(k)=k^n$ for $n=1$ (short-dashed line) and $n=2$ (long-dashed line).
The linear behavior observed here implies density of states
increasing as $(...)^N$ and a T=0 residual entropy.
}
\label{fig:nb-singlet}
\end{figure}

This phase was first observed in the Heisenberg model on the kagom\'e 
lattice~\cite{ce92,le93,web98,smlbpwe00}.
Some possible physical realizations are $SrCrGaO$ 
or various jarosites which have kagom\'e planes of
magnetic ions\cite{r94,wills-98}.
Fig~\ref{fig:1}d shows the spectrum of the $N=36$ sample 
of the kagom\'e antiferromagnet.
Two features, confirmed by finite size scaling, are visible.
First, a finite, although small, spin-gap $\Delta$.
So $SU(2)$ spin symmetry remains unbroken and $\chi$ is exponentially activated.
Second, a gapless continuum of singlet levels filling the spin-gap.
Futhermore, the number of singlet levels in the spin-gap increases
exponentially with the size $N$ of the sample
as shown in Fig~\ref{fig:nb-singlet}.
All the correlations (spin-spin, dimer-dimer, chiral-chiral, ...)
are very small and short range~\cite{ce92,le93}.

\begin{figure}[t]
 \vspace{-11pc}
 \hspace{2pc}
\epsfxsize=22pc
\epsfbox{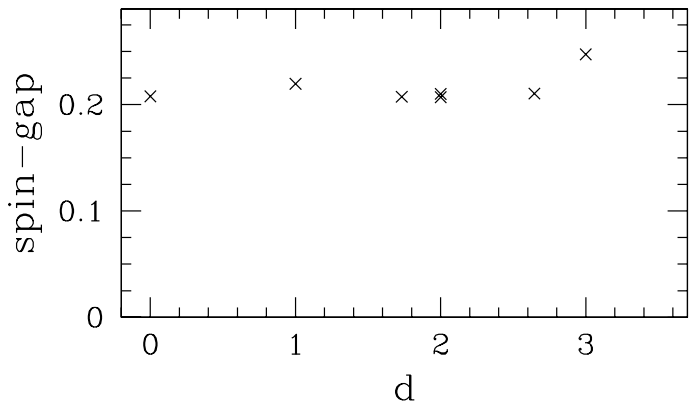}
\caption{Spin-gap of kagom\'e $N=24$ sample depletted of two sites
vs the euclidian distance $d$ between the two empty sites.
The $d=0$ value is the spin-gap of the undepleted sample.
}
\label{fig:gap_vs_hole_dist}
\end{figure}

The understanding of the low lying continuum of singlets is still
incomplete but some progresses have been done
in the last years\cite{sc-progress,m98,mm01,gsf-01}. 
Recent ED calculations\cite{sc-progress}
on a kagom\'e $N=24$ sample depletted of two sites 
suggest that the elementary excitations likely consist of deconfined spinons:
as shown in Fig~\ref{fig:gap_vs_hole_dist}, the spin-gap does not
depend neither on the presence of the two nonmagnetic holes nor on their
separation  which
suggest that spin-1/2 excitations are essentially free
(unconfined)\cite{sachdev-vojta-00}.
The low lying singlets could be related to a family of short
range RVB dimer coverings of the lattice~\cite{m98,mm01}.
On the other hand, they cannot be described as 
Goldstone modes associated to a quasi LRO of dimers.
Assuming non-interacting modes with a dispersion law $\epsilon(k)=k^n$,
it is easy to obtain the internal energy and entropy versus temperature 
and thus the number of states below a given energy. 
The logarithm of this number of states scales as
$N^{\frac{n}{n+2}}$. 
Fixing the unknown constants from the measured value 
of the number of singlets in the spin-gap for $N=12$,
we see in Fig~\ref{fig:nb-singlet}
that the mode description is an order of magnitude off at $N=36$.
The  exponential increase in the number of eigenlevels  in the
finite energy interval below the spin-gap suggests 
a finite entropy per spin at T=0.
The same exponential increase in the number of eigenlevels
in each spin sector can be at the origin of the anomalous 
glassy behavior "without chemical disorder" 
seen in several kagom\'e compounds\cite{wdvhc00},
on which dynamical mean-field theory approaches
might shed light\cite{gsf-01}.
A low lying continuum of singlets is consistent with the 
observation\cite{rhw00} 
that the low temperature specific heat of $SrCrGaO$ is essentially insensitive
to large magnetic fields\cite{smlbpwe00}.
This picture is also compatible with the
muons experiments of Uemura {\it et al}\cite{ukkll94} and the elastic spin
diffusion measurements of Lee {\it et al}\cite{lbar96}. 

Type II Spin-Liquid behavior might also occur
in 3d, in pyrochlore compounds\cite{gardner-99} %\cite{canals-lacroix-98}
and gadolinium gallium garnet\cite{tsui-01}. 
Likewise, 2d systems of coupled 1d chains might 
exhibit Spin-Liquid like behavior\cite{coldea-01,starykh-01}.
Lately, neutron scattering evidence for deconfined spinons on an anisotropic
triangular lattice has been found\cite{coldea-01}.

For a long time, it was widely believed 
that type II Spin-Liquid behavior was
associated to special lattices like the kagom\'e or pyrochlore lattices,
which at
the classical level display an infinite local ground-state degeneracy.
We have now several counterexamples.
The checkerboard lattice which is a 2d analog of the pyrochlore lattice
and exhibit a large similar degeneracy at the classical level,
was found to display VBC LRO\cite{fmsl01}.
A kagom\'e-like spectra is observed on the
triangular lattice when the 3-sublattice N\'eel order is destabilized by a
frustrating 4-spin exchange\cite{lmsl00} and we have no evidence of an
infinite local degeneracy at this point.
We now conjecture that the "kagom\'e-like" phase is one
of the possible scenario after destabilization  a non-collinear LRO.
At this point we have not made any difference between the criteria of 
appearance of the two Spin-Liquids.
Attempting, to go further, one could speculate that 
type I  Spin-Liquid behavior might originate from resonating entities
including near neighbor triplet pairs assembled in singlet entities
of at least 4 spins, whereas type II Spin-Liquid involve essentially
near neighbor singlet pairing.
A last hint comes from the study of the spin-$S=1$ kagom\'e
magnet\cite{k_hida00}, for sizes up to $N=18$, which point to a rather
large spin-gap void of singlets.
It is in fact  a common general idea that,
due to topological reasons,
half-integer and and integer spin systems could differ.
The difference between spin-1/2 and spin-1 kagom\'e
spectra might possibly originate from this topological property.

\end{document}